# LENS OR RESONATOR ? - ELECTROMAGNETIC BEHAVIOR OF AN EXTENDED HEMIELLIPTIC LENS FOR A SUB-MM WAVE RECEIVER

**Artem V. Boriskin[1,3], Alexander I. Nosich[1,2]**
**Svetlana V. Boriskina[2], Trevor M. Benson[2]**
**Phillip Sewell[2], and Ayhan Altintas[3]**
[1] Department of Computational Electromagnetics
Institute of Radio-Physics and Electronics
National Academy of Sciences
Kharkov 61085, Ukraine
[2] George Green Institute for Electromagnetic Research
University of Nottingham, NG7 2RD, UK
[3] Department of Electrical and Electronics Engineering
Bilkent University, 06533 Ankara, Turkey

**ABSTRACT:** *The behavior of a 2-D model of an extended hemielliptic silicon lens of a size typical for THz applications is accurately studied for the case of a plane E-wave illumination. The full-wave analysis of the scattering problem is based on the Muller's boundary integral-equations (MBIE) that are uniquely solvable. A Galerkin discretization scheme with a trigonometric basis leads to a very efficient numerical algorithm. Numerical results related to the focusability of the lens versus its rear-side extension and the angle of the plane-wave incidence, as well as near-field profiles, demonstrate strong resonances. Such effects can change the principles of optimal design of lens-based receivers.*

## I. INTRODUCTION

Planar slot or strip elements combined with dielectric lenses are attractive building blocks for mm and sub-mm wave receivers [1]-[5]. The attention they have been attracting recently is due to their capability for compact integration with other electronic components such as detecting diodes, local oscillators and mixers. Furthermore, they provide better efficiency than other types of antennas printed on homogeneous substrates. The elliptical shape of the lens is borrowed from optics to provide focusing properties if its eccentricity is selected properly. On the other hand, the lens interface gives rise to reflections inside a realistic lens that may significantly affect the input impedance and sensitivity. This aspect, which has not been properly investigated in the literature, is a critical point in the overall design of lens antennas [2], [5]. Various analytical techniques used to simulate dielectric lenses have been commonly based on ray tracing and neglecting the lens size and curvature, and hence have failed to characterize the internal resonances. The validity of such approximations can be questionable in many cases as the actual size of the lens is several wavelengths [1]-[6]. In fact, the behavior of such a lens should clearly display both ray-like and resonance-mode features. On the other hand, FDTD-based solvers are time and memory consuming, require accurate treatment of the back-reflections from the edge of the computational window, and may suffer from a loss of accuracy at resonant frequencies. All the above shows that the true electromagnetic behavior of dielectric lenses is still far from clear.

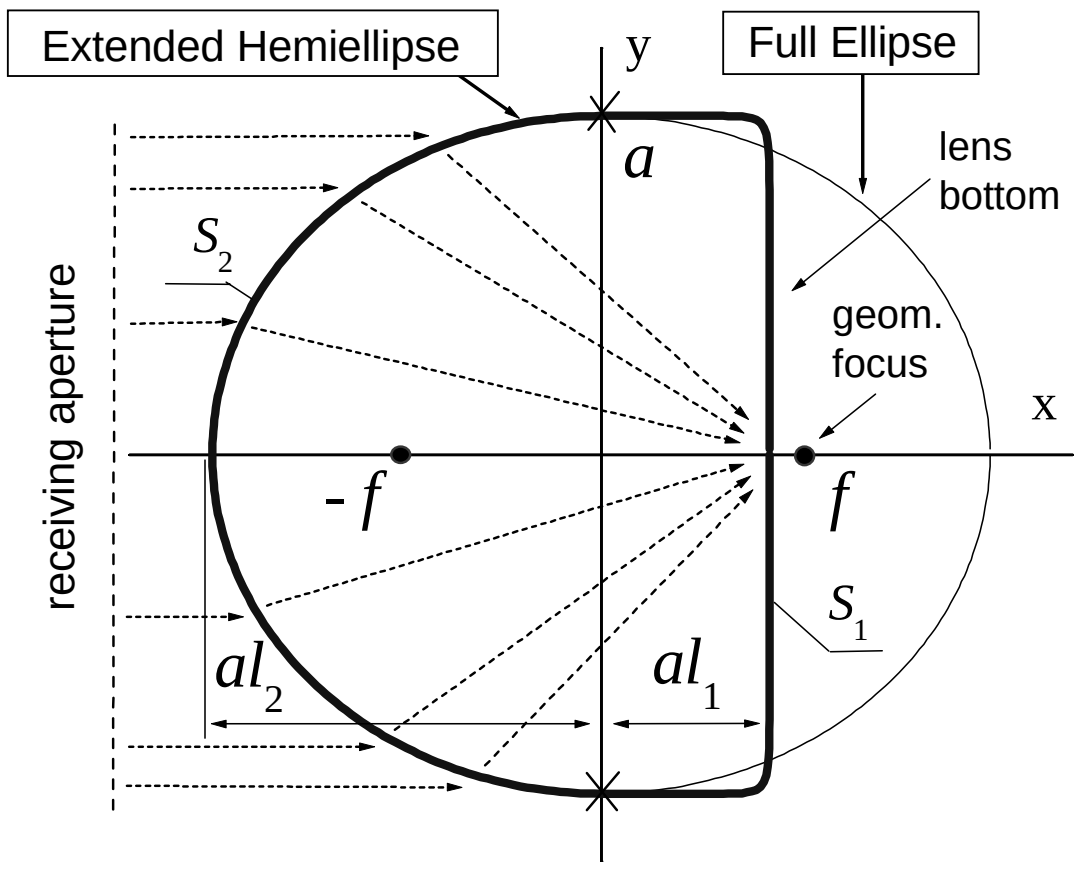

**Figure 1** Geometry and notations for an extended hemielliptic lens. The foci of the full ellipse are marked with black dots.

In this paper, we analyze extended hemielliptic lenses in the receiving mode. If the eccentricity of the ellipse relates to the lens dielectric constant as

$$e = 1/\sqrt{\varepsilon}\,, \tag{1}$$

then geometrical optics (GO) predicts that all the rays of a parallel family that impinge on the front surface of the lens are collected at the rear focal point (Fig.1). In applications, it is important to know the actual shape of the focal domain for various lens sizes and various angles of incidence of the plane wave.

## II. OUTLINE OF SOLUTION

A brief outline of the MBIE method is as follows. In two-dimensions, the studied problem can be formulated as plane-wave scattering by a dielectric cylinder with a smooth contour of cross-section, whose frontal part follows (1). The application of Green's identity enables one to express the fields inside and outside the lens in terms of the limiting values of the fields and their normal derivatives on the cylinder contour. By applying transmission-type boundary conditions, a set of Fredholm second-kind IEs is obtained, known as Muller's boundary IEs [7]. These can be discretized using the collocation method [8]. However, we enhance the convergence of the discretized solutions by applying the Galerkin method with entire-domain angular exponents (also known as trigonometric polynomials) as the expansion functions. To treat the singularities in the Green's functions and their derivatives in the discretization scheme, we use an analytical integration of the canonical circular-cylinder counterpart terms. Such a procedure enables us to exploit the fact that the trigonometric polynomials are orthogonal eigen-functions of the canonical-shape operators.

This procedure results in a Fredholm second-kind infinite-matrix equation having favorable features, with the right-hand-part terms and matrix elements obtained as Fourier-expansion coefficients of twice-continuous functions. These can be economically computed using the Fast Fourier Transform (FFT) and Double FFT algorithms, respectively. The algorithm developed guarantees point-wise convergence of the numerical solution, i.e., the possibility of minimizing the error to machine precision by solving progressively larger matrices. Note that this convergence is a universal one - the required accuracy can be achieved for an arbitrary set of lens parameters, e.g., for any value of the contrast between the lens material and the background medium. It is also free from the inaccuracies near to the sharp natural resonances that are intrinsic to conventional numerical approximations [9], as well as from the well-known defect of "numerical resonances" occurring if the fields are derived from single or double layer potentials [10]. More details of the algorithm's properties can be found in [11].

### III. NUMERICAL RESULTS

We shall consider the cross-section of the extended hemielliptic lens bounded by a curve $S$, which is twice-differentiable at every point and consists of two parts, $S_1$ and $S_2$ , smoothly joined together (Fig.1). Here, $S_2$ is one half of the ellipse whose eccentricity is assumed in accordance with (1), and $S_1$ is one half of a so-called "super-ellipse" that approximates a rectangle with rounded corners. Both curves can be characterized by the same equation:

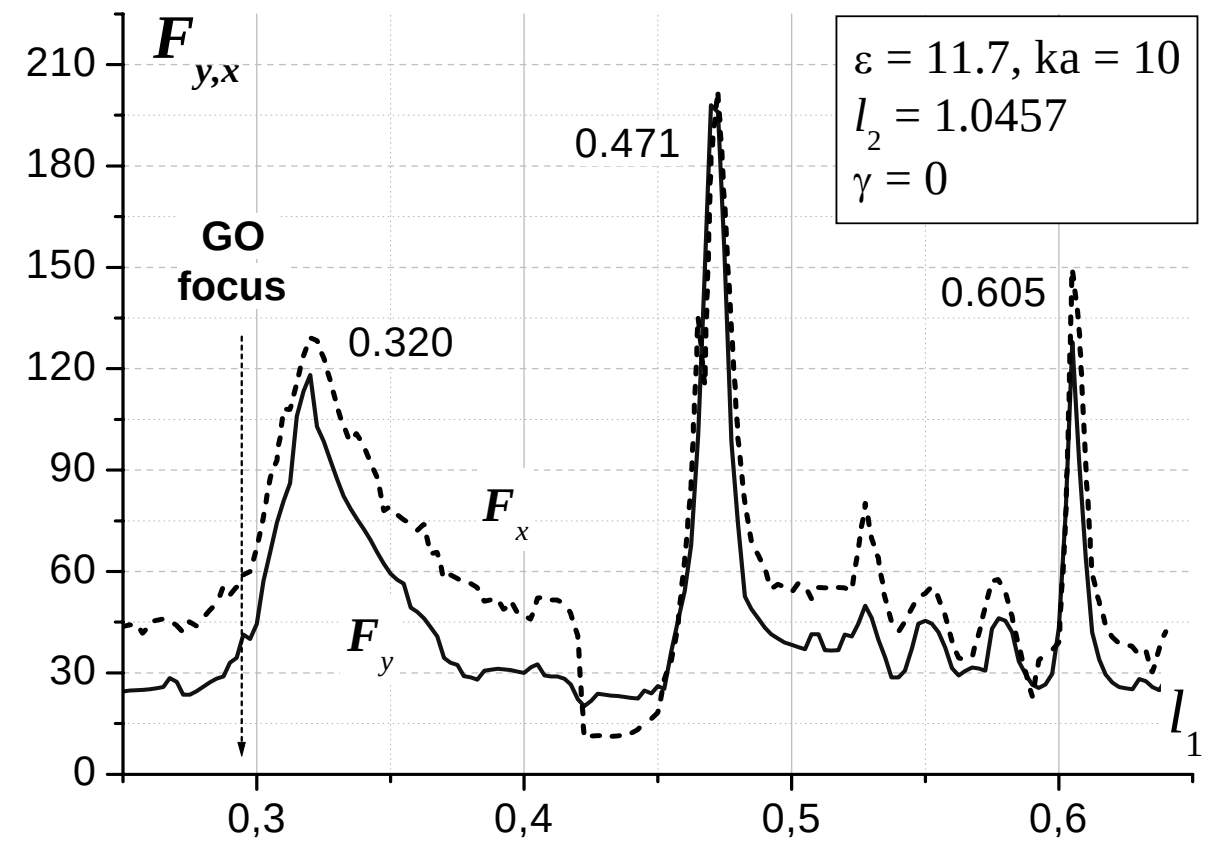

**Figure 2.** Focusability of the extended hemielliptic lens versus the relative lens extension parameter.

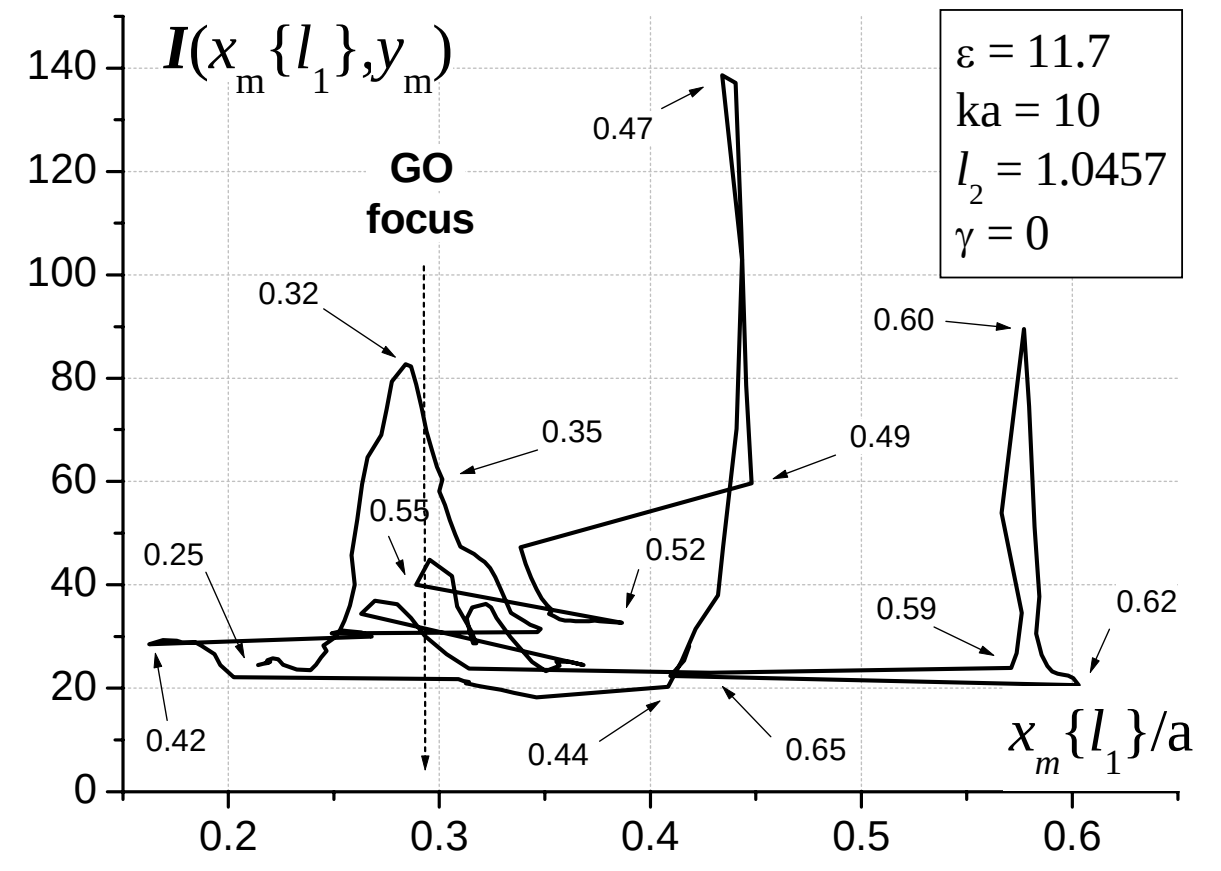

**Figure 3.** The $x$-coordinate and the peak field intensity for the extended hemielliptic lens versus lens extension parameter. Characteristic values of $l_1$ are indicated by arrows.

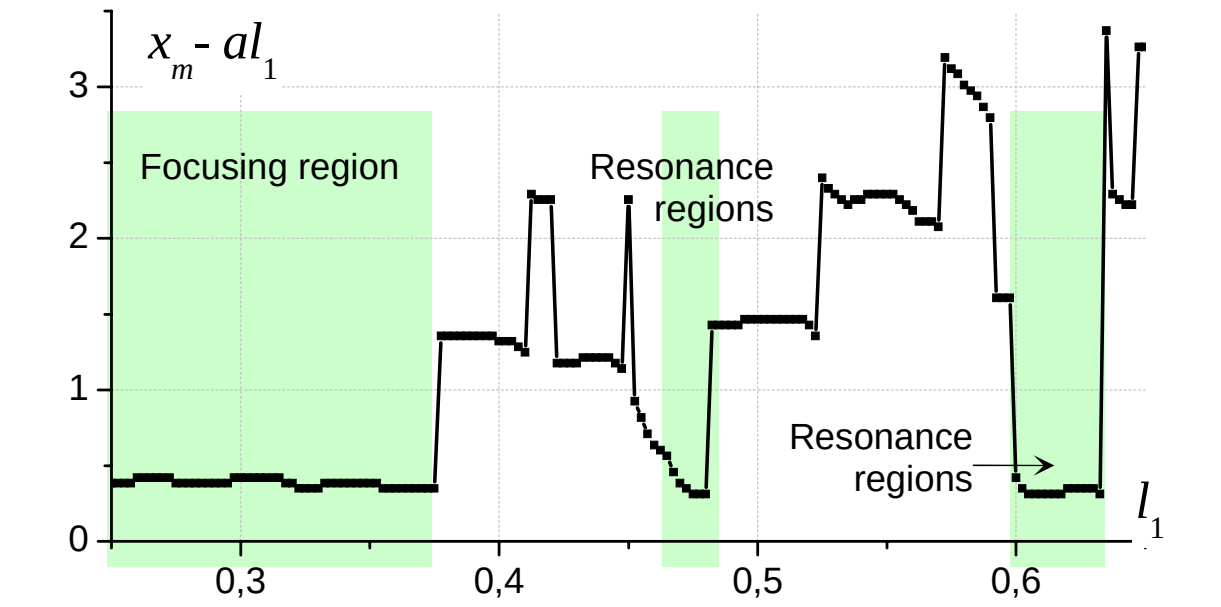

**Figure 4.** Normalized distance between the peak field intensity point $x$-coordinate and the lens bottom versus the lens extension parameter.

$$(x/l_j a)^{2\nu_j} + (y/a)^{2\nu_j} = 1, \qquad j = 1,2 \tag{2}$$

In the specific computations presented here, we shall use a silicon lens ($\varepsilon$=11.7) whose size is approximately 3 free-space wavelengths (*ka*=10), and the contour parameters are $\nu_1$=10 and $\nu_2$=1, respectively.

The concentration of the field along the *x*- and *y*-axes can be quantified in terms of the "focusability" defined as

$$F_x = \frac{aI(x_m, y_m)}{|x_m - x_{0.5}|}, \quad F_y = \frac{aI(x_m, y_m)}{|y_m - y_{0.5}|}, \tag{3}$$

where $(x_m, y_m)$ are the coordinates of the maximum field intensity ($I = |E_z|^2$) inside the lens, and $(x_{0.5}, y_{0.5})$ are the coordinates where the field intensity is a half of the maximum value.

Proceeding from GO, one usually considers that the focusability of the true elliptic lens is preserved even if one removes its rear part behind the GO focus. For a wavelength-scale lens this is not true.

This claim is supported by Fig. 2, which shows the focusability of the lens versus its relative bottom extension parameter, $l_1$. One can see that it increases monotonically with $l_1$ approaching and even exceeding the value corresponding to the full ellipse focus location (marked with a dashed line). In Fig. 3, one can see the movement of the point with the highest field intensity along the *x*-axis and the field intensity peak value versus the same parameter (figures mark the values of $l_1$ at the points indicated with arrows).

Analysis of the graphs shows that there are two mechanisms that explain the internal field enhancement. One is the focusing that causes a wide peak in the focusability of Fig. 2 near to the value of $l_1$ that corresponds to the GO focus location, $l_1$=0.2924. Another is a series of internal resonances, which results in a sequence of sharp features, in both the focusability and the peak field intensity. Note that this behavior is not predicted by GO. Another fact that can be of interest for antenna designers is that the position of the highest-intensity spot migrates in a complicated manner between the bottom of the lens and the location of the GO focus (Fig. 3). One can see that the spot approaches the lens bottom if either it is cut near to the GO focus or if it is cut to exploit an internal resonance (Fig. 4). Note that the peak intensity for the extended hemielliptic lens tuned to an internal resonance (here and after referred to as a *resonant lens*) can reach over 300% in comparison to the commonly used lens extended to the GO focus location (here and after referred to as a *quasi-optical lens*).

The near-field profiles for the resonant values of $l_1$ revealed in Fig. 2 are shown in the bottom of Fig. 5. These resonances are closely related to the asymptotic GO resonances known as "bouncing-ray" or "billiard-type" ones. Here they have specific triangular profiles and can be classified with two indices, which count the field variations along the lens bottom (*m*) and a side of the triangle (*n*). Two neighboring resonances in $l_1$ have *n* and $n+1$ variations, and the distance between them corresponds to half a wavelength in dielectric along the triangle side.

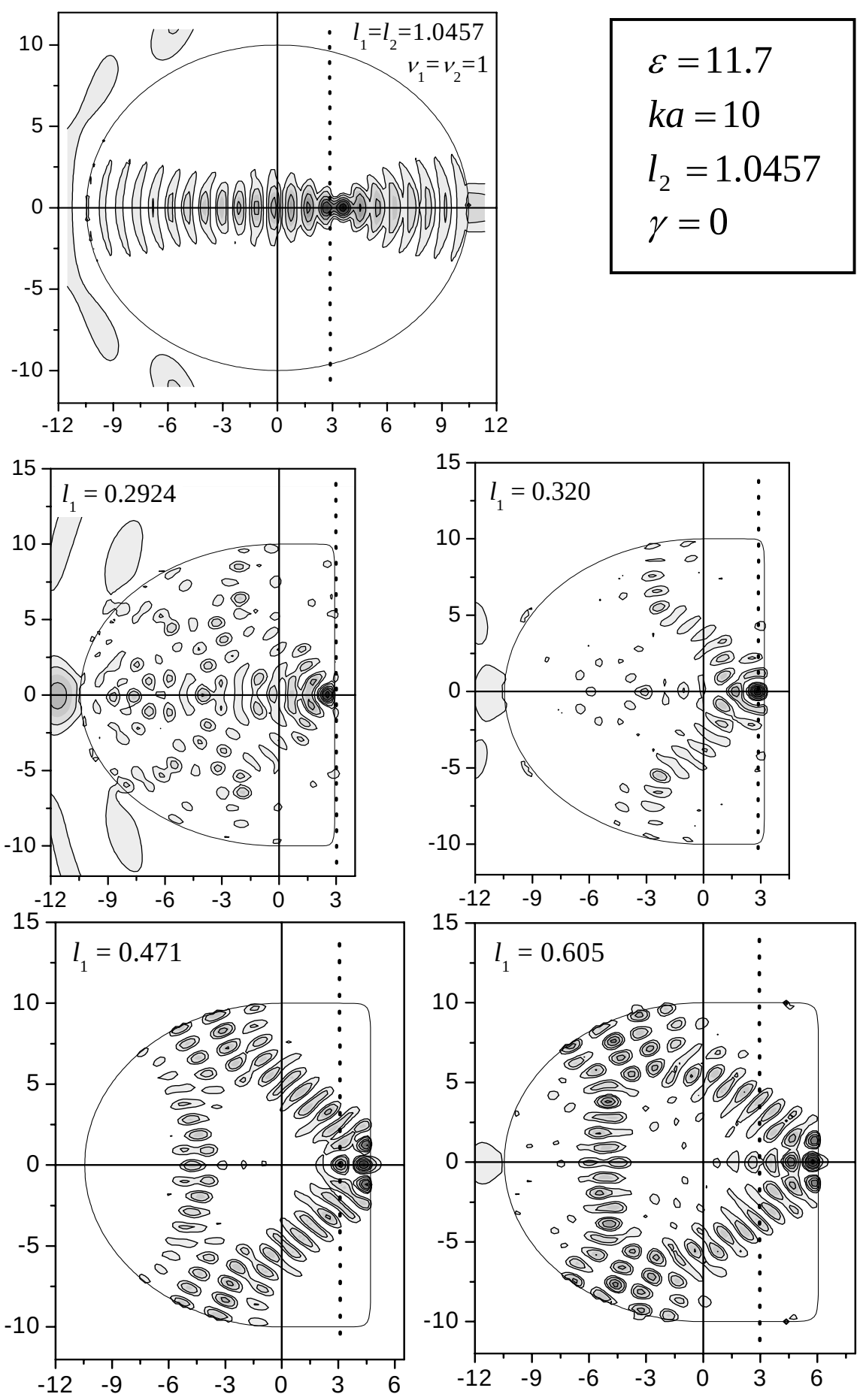

**Figure 5.** Near-field intensity portraits for the full-elliptic and extended hemielliptic silicon lenses with different extensions. Dotted lines mark the location of the rear focus of full ellipse.

Therefore, this accurate study has shown that the resonances are very important to understand the behavior of a wavelength-scale lens. Furthermore, replacement of the full ellipse by an extended hemiellipse leads to a dramatic change in the field distribution inside the lens. Nevertheless, a proper choice of the lens rear-side extension may lead to a considerable improvement in the lens' ability to produce field spots with large intensity and hence boost receiver sensitivity.

Fig. 6 shows the field intensity along the flat side for a quasi-optical lens and a resonant lens tuned to the resonance at $l_1 = 0.471$, for several angles of plane wave incidence. It can be seen that the quasi-optical lens has a single spot of high field intensity located at the flat side. If the incidence angle $\gamma$ of the plane wave (counted from the *x*-axis) varies, the main field maximum moves along the lens flat side and broadens. At $\gamma$=20$^0$, the field intensity in the main maximum falls to a half of its value with normal incidence, while the secondary maxima rise to nearly the same level.

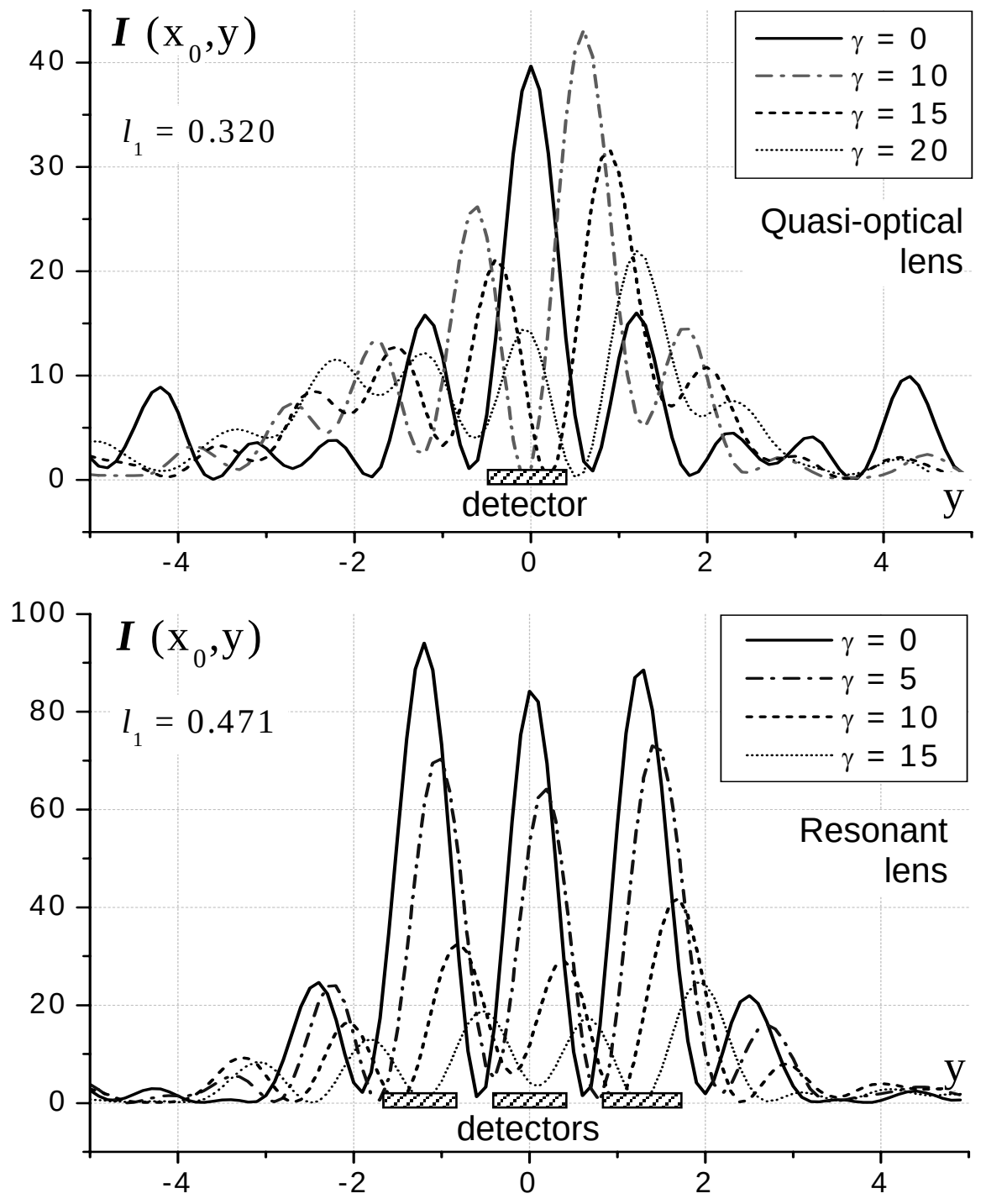

**Figure 6.** Field intensity along the flat sides of two differently extended lenses for a varying angle of the plane wave incidence. (see insets) $x_0 = al_1, \varepsilon = 11.7, ka = 10,\ l_2 = 1.0457$.

In contrast to this, for the resonant lens three spots of the same peak intensity are observed. Each peak is twice as large as that of the quasi-optical lens. As the incidence angle of the plane wave varies, these high-field spots maintain fixed locations and the maximum intensity drops by half if $\gamma = 8^0$, while the side-spot peaks remain at 40% of the peak intensity for normal incidence.

This analysis shows that a silicon lens tuned to an internal resonance has certain advantages, despite losing its wideband nature. Particularly attractive are the stability of the resonant field with respect to the angle of arrival of incident wave and the fact that several times greater values of the peak intensity can be achieved. Narrow-band receivers exploiting these effects may potentially have improved sensitivity and scanning performance.

## IV. CONCLUSIONS

The electromagnetic behavior of an extended hemielliptic lens used in mm and sub-mm wave receivers has been studied in a 2-D formulation. An efficient and accurate numerical method based on the MBIE technique has been applied to the solution of the plane wave scattering by a smooth dielectric cylinder whose contour of cross-section follows the shape of such a lens.

Our numerical results demonstrate effects that cannot be predicted with GO or physical optics approximations. The most important feature revealed by the accurate analysis is that resonances may play a dominant role in the wavelength-scale lens behavior. We have also provided particular numerical evidence that advanced character-ristics of the lens-coupled receivers can be achieved, although in a narrow band, by exploiting a resonance.

Therefore, the answer to the question put in the title is that a typical THz lens antenna made of silicon is rather a *dielectric resonator* than a lens.

## ACKNOWLEDGEMENT

This work has been supported by the UK EPSRC under grants reference GR/R65213/01 and GR/S60693/01(P) and by the TUBITAK grant No. EEEAG-103E037.